\documentstyle[12pt,epsfig]{article}

\def\Journal#1#2#3#4{{#1} {\bf #2}, #3 (#4)}

\def\PLB{{\em Phys. Lett.} B}

\def\PRP{\em Phys. Rep.}
\def\IJMP{{\em Int. J. Mod. Phys.} A}
\def\PRD{{\em Phys. Rev.} D}

\renewcommand{\thefootnote}{\fnsymbol{footnote}}

\begin{document}

\newpage
\setcounter{page}{0}

\begin{titlepage}
\begin{flushright}
\hfill{YUMS 98-3}\\
\hfill{\today}
\end{flushright}
\vspace{2.0cm}

\begin{center}
{\large\bf CHARGINO PAIR PRODUCTION AT CERN LEP II\footnote{Presented at the
           APCTP workshop: ``Pacific Particle Physics Phenomenology"
           (Oct. 31 - Nov. 2, 1997).}}
\end{center}
\vskip 1.3cm
\begin{center}
S.Y. Choi\\
\vskip 0.3cm
{\em Department of Physics, Yonsei University, Seoul 120-749, Korea}
\end{center}

\vskip 4cm

\setcounter{footnote}{0}
\begin{abstract}
Charginos are expected to be the lightest observable supersymmetric 
particles in many of the supersymmetric models. In the scenario that lighter
charginios are pair-produced at CERN LEP II, we present a straightforward 
procedure for determining the SUSY parameters, $\tan\beta$, $M_2$, $\mu$, and
the electron sneutrino mass $m_{\tilde{\nu}}$ up to a four-fold discrete
ambiguity, although a large number of unknown SUSY parameters are involved
in chargino decays.
\end{abstract}
\end{titlepage}

\newpage
\renewcommand{\thefootnote}{\alph{footnote}}

One of the main goals of the CERN $e^+e^-$ collider at LEP II\cite{LEP2_Physics}
is to search for signs of weak-scale supersymmetry (SUSY)\cite{Haber_Kane}. 
Among many SUSY particles that might be found, the chargino, a mixture of the 
$W$-ino and charged Higgsinos, is of particular interest. 
From the theoretical point of view, charginos are expected to be lighter than 
gluinos and most sfermions,  and for this reason, chargino searches have been 
well studied. The current lower bound on the lighter chargino mass is about 
90 GeV\cite{ALEPH}. Chargino discovery studies have shown that
the discovery reach will extend nearly to the LEP II kinematical limit. 
Moreover, if kinematically accessible, charginos have a 
large cross section throughout SUSY parameter space and produce a clean signal
in certain decay modes.  As the chargino pair production cross 
section rises rapidly above threshold, each step in collider energy holds
the promise not only of chargino discovery, but also of detailed SUSY studies
from chargino events. 

The purpose of the present work is to estimate the capability of LEP II
to determine the parameters of SUSY in lighter chargino pair production in
the Minimal Supersymmetric Standard Model (MSSM). This issue has been studied
by Leike\cite{Leike}, Diaz and King\cite{Diaz_King} and recently by Feng
and Strassler\cite{Feng_Strassler}. Our approach here is 
to employ the most dominant chargino decay modes to extract the full possible
information on chargino polarizations, while avoiding  theoretical assumptions 
at high energy scales, and exploiting the fact that the chargino-pair 
production process depends on only a small subset of the SUSY parameters.
In this case,  although chargino decays 
are very complicated with many diagrams involved, certain final-state angular 
correlations, which are experimentally identifiable, allow us to construct 
three additional observables besides the production cross section so that all
the SUSY parameters relevant to chargino pair production are determined up to 
a discrete ambiguity.

The MSSM includes the usual matter superfields and two Higgs doublet 
superfields $\hat{H}_1$ and $\hat{H}_2$, which give masses to the 
isospin $-\frac{1}{2}$ and $+\frac{1}{2}$ fields, respectively. 
These two superfields are coupled in the
superpotential through the term $-\mu\epsilon_{ij}\hat{H}^i_1\hat{H}^j_2$,
where $\mu$ is the supersymmetric Higgs boson mass parameter. The ratio of the 
two Higgs scalar vacuum expectation values is defined to be 
$\tan\beta\equiv\langle H^0_2\rangle/\langle H^0_1\rangle$.
There are two chargino mass eigenstates that result from the mixing of the
electroweak gauginos $\tilde{W}^\pm$ with the Higgsinos due to the spontaneous
electroweak symmetry breaking. The chargino mass term is 
\begin{eqnarray}
-{\cal L}_m=\left(\bar{\tilde{W}^-_R} \bar{\tilde{H}^-_{2R}}\right)
            \left(\begin{array}{cc}
                M_2                &      \sqrt{2}m_W\cos\beta  \\
             \sqrt{2}m_W\sin\beta  &             \mu   
                  \end{array}\right)
            \left(\begin{array}{c}
                  \tilde{W}^-_L  \\
                  \tilde{H}^-_{1L}
                  \end{array}\right) + {\rm h.c.}.
\end{eqnarray}
We assume that the gaugino masses $M_i$ ($i=1,2$) and the parameters $\mu$ and 
$\tan\beta$ are real so that $CP$ violation plays no role in 
chargino events. Then the $2\times 2$ complex chargino mass matrix can 
be diagonalized by two orthogonal matrices $O_L$ and $O_R$ defined by two
rotation angles $\phi_L$ and $\phi_R$, respectively, which
appear at the couplings between the gauge boson $Z$ and the charginos.

In general, studying MSSM is complicated because of many free 
parameters so that we make the following assumptions;
(i) $R$ parity is conserved, (ii) the lightest supersymmetric particle (LSP) 
is the lightest neutralino $\tilde{\chi}^0_1$, (iii) sfermions have masses 
beyond the LEP II kinematical limit, and (iv) the intergenerational mixing 
in the sfermion and quark sectors is small and may be neglected.
Under our assumptions, for almost all values of parameters, charginos decay to 
three-body final states consisting of an LSP and either two quarks or two 
leptons at the LEP II c.m. energy.

One typical observable which is crucial to our analysis is the chargino 
mass $m_{\tilde{\chi}^\pm_1}$. The mass $m_{\tilde{\chi}^\pm_1}$ 
along with the LSP mass  $m_{\tilde{\chi}^0_1}$ can be measure by
the dijet energy distribution in hadronic chargino decays\footnote{An 
alternate determination of $m_{\tilde{\chi}^\pm_1}$ can be provided by
an energy scan at the chargino production threshold\cite{Leike}.}. 
The end points of the dijet energy $E_{jj}$ and mass spectra are completely 
determined by $m_{\tilde{\chi}^\pm_1}$ and $m_{\tilde{\chi}^0_1}$ with 
the maximal and minimal dijet energies, $E_{max}$ and $E_{min}$, as
\begin{eqnarray}
m_{\tilde{\chi}^\pm_1}=\frac{\sqrt{E_{max}E_{min}}}{E_{max}+E_{min}}
                       \sqrt{s},\qquad
m_{\tilde{\chi}^0_1}=\sqrt{1-2\frac{(E_{max}+E_{min})}{\sqrt{s}}}.
\end{eqnarray}
The distribution of the invariant mass $m_{jj}$ is between zero and 
$m_{\tilde{\chi}^\pm_1}-m_{\tilde{\chi}^0_1}$. If at least two of the 
three end points are sufficiently sharp to be well measured, they
can be used to precisely determine $m_{\tilde{\chi}^\pm_1}$ and
$m_{\tilde{\chi}^0_1}$. 

The amplitude for producing charginos which decay to a certain final state
does not factorize into production and decay amplitudes due to the fact that
charginos are spin-$\frac{1}{2}$ objects so that the angular correlations
of chargino decay products are affected by the underlying structures of 
the production processes. In the present work, we will consider the production 
process $e^+e^- \rightarrow \tilde{\chi}^-_1\tilde{\chi}^+_1$ followed by the 
decays $\tilde{\chi}^-_1\rightarrow \tilde{\chi}^0_1 f_1\bar{f}_2$
and $\tilde{\chi}^+_1\rightarrow \tilde{\chi}^0_1 f_3\bar{f}_4$ where
$f,f' (g,g')$ are for quarks and/or leptons, and then 
we can write the amplitude for the sequential process in the narrow 
width approximation as a multiplication of the production helicity
amplitudes $T_{\sigma;\lambda\bar{\lambda}}$ and two decay helicity
amplitudes ${\cal D}_\lambda$ and $\bar{\cal D}_{\bar{\lambda}}$ for the
negative and positive charginos, where $\sigma=\pm$ is the electron helicity, 
$\lambda,\bar{\lambda}$ are the negative and positive chargino helicities.
With the fermion masses neglected, 
$\tilde{\chi}^-_1\rightarrow \tilde{\chi}^0_1 f_1\bar{f}_2$ includes
two sfermion-exchange diagrams and one $W$-boson exchange, and the decay
amplitudes depend on only the invariant mass of two final fermions
because the effects of the sfermions can be well approximated by 
point propagators\cite{Feng_Strassler} under the assumption that sfermions
have masses beyond the LEP II kinematical limit.

In order to evaluate the chargino decay helicity amplitudes, we take the
chargino rest frames and introduce the angular variables ($\theta^*,\phi^*$)
and ($\bar{\theta}^*,\bar{\phi}^*$) for the two-fermion systems in the 
negative and positive chargino decays, respectively. 
Maintaining only the angular dependence and integrating the decay distributions
over the invariant masses $q^2_W$ and $\bar{q}^2_W$, we find that when the
$f_3\bar{f}_4$ system is charge-conjugate to the $f_1\bar{f}_2$ system,
the decays for polarized negative and positive charginos are 
described by the decay density matrices
\begin{eqnarray}
&& \rho_{\lambda\lambda^\prime}
   =\frac{{\cal D}_\lambda {\cal D}^*_{\lambda^\prime}}{
            \sum_{\lambda}{\cal D}_\lambda {\cal D}^*_{\lambda}}
   =\frac{1}{2}\left(\begin{array}{cc}
         1+\kappa\cos\theta^*         &   \kappa\sin\theta^*{\rm e}^{i\phi^*} \\
 \kappa\sin\theta^*{\rm e}^{-i\phi^*} &        1-\kappa\cos\theta^*
                     \end{array}\right),\nonumber\\
&& \bar{\rho}_{\bar{\lambda}\bar{\lambda}^\prime}
   =\frac{\bar{\cal D}_{\bar{\lambda}}\bar{\cal D}^*_{\bar{\lambda}^\prime}}{
            \sum_{\bar{\lambda}}\bar{\cal D}_{\bar{\lambda}}
                                \bar{\cal D}^*_{\bar{\lambda}}}
   =\frac{1}{2}\left(\begin{array}{cc}
 1-\kappa\cos\bar{\theta}^* & \kappa\sin\bar{\theta}^*{\rm e}^{i\bar{\phi}^*} \\
 \kappa\sin\bar{\theta}^*{\rm e}^{-i\bar{\phi}^*} & 1-\kappa\cos\bar{\theta}^*
                     \end{array}\right),
\end{eqnarray}
respectively. The parameter $\kappa$, which determines the angular dependence 
of the decay distributions, is a function of chargino and neutralino mixing 
parameters, chargino and neutralino masses, and sfermion masses so that there
is a very wide variety in estimating the parameter value. 
Therefore, we consider $\kappa$ as a phenomenological parameter 
in our analysis, which is not fixed. 

Combining the production and decay amplitudes yields a five-dimensional 
differential cross section consisting of a kinematical factor and
an angular-dependent part
$\Sigma(\Theta;\theta^*,\phi^*;\bar{\theta}^*,\bar{\phi}^*)$ where
$\Theta$ is the $\tilde{\chi}^-_1$ production angle.
The angular dependence $\Sigma$ for the case that two two-fermion systems are
charge-conjugate to each other is decomposed into eight independent
parts:
\begin{eqnarray}
\Sigma&=&\Sigma_{\rm unpol}+(\cos\theta^*+\cos\bar{\theta}^*)\kappa P
          +\cos\theta^*\cos\bar{\theta}^*\kappa^2 Q\nonumber\\
       && +(\sin\bar{\theta}^*\cos\bar{\phi}^*-\sin\theta^*\cos\phi^*)\kappa W
            \nonumber\\
       && +(\cos\theta^*\sin\bar{\theta}^*\cos\bar{\phi}^*
           -\cos\bar{\theta}^*\sin\theta^*\cos\phi^*)\kappa^2 X \nonumber\\
       && +\sin\theta^*\sin\bar{\theta}^*\cos(\phi^*+\bar{\phi}^*)
           \kappa^2 Y
          +\sin\theta^*\sin\bar{\theta}^*\cos(\phi^*-\bar{\phi}^*)
           \kappa^2 Z,
\end{eqnarray}
where the eight distribution functions describing chargino production are 
defined in terms of the production helicity amplitudes
$T_{\sigma;\lambda\bar{\lambda}}$ (see Ref.~7 for the 
definition of the eight distribution functions.).

Certainly not all of the distribution functions are experimentally 
measurable. As for the reconstruction problems in chargino production
and decays we have to take into account the following aspects.
First of all,  because the LSP is not detected, there exists at least a 
two-fold discrete ambiguity in determining the scattering angle $\Theta$
for the hadronic decay modes of the negative and positive charginos.
So, we consider the distributions integrated over the scattering 
angle $\Theta$. Nevertheless, $\cos\theta^*$, $\cos\bar{\theta}^*$
and $\sin\theta^*\sin\bar{\theta}^*\cos(\phi^*-\bar{\phi}^*)$ are
measurable experimentally through the relations
\begin{eqnarray}
&&\cos\theta^*=\frac{1}{\beta q^*}\left(\frac{E}{\gamma}-E^*\right),\qquad
    \cos\bar{\theta}^*=\frac{1}{\beta \bar{q}^*}
              \left(\frac{\bar{E}}{\gamma}-\bar{E}^*\right),\nonumber\\
&&\sin\theta^*\sin\bar{\theta}^*\cos(\phi^*-\bar{\phi}^*)=
     \cos\vartheta
    -\frac{\left(E-\frac{E^*}{\gamma}\right)
     \left(\bar{E}-\frac{\bar{E}^*}{\gamma}\right)}{\beta^2 q\bar{q}},
\end{eqnarray}
where ($E,q$) and ($\bar{E},\bar{q}$) are the energy and absolute
momentum of two two-fermion systems in the laboratory frame, 
$\gamma=\sqrt{s}/{2m_{\tilde{\chi}^\pm_1}}$,  the angle $\vartheta$ is the 
angle between the two hadronic systems, and the superscript $*$ denotes
energy and momenta in the chargino rest frames.
Secondly,  $P$ is parity-odd and $C$-odd so that its determination
requires charge identification of charginos. This can be accomplished
by the mixed mode of the leptonic and hadronic decays of the negative and 
positive charginos where only the lepton charge is needed to be
identified. On the other hand, the $C$-even $Q$ and $Z$ do not
require charge identification so that the dominant hadronic decay
mode for both negative and positive charginos can be used. 
Therefore, if the total cross section $\sigma_{tot}$ and the total leptonic 
and hadronic branching ratios of the chargino decays are experimentally 
determined, we can have three additional independent observables
integrated over the scattering angle $\Theta$;
$\kappa\langle P\rangle$, $\kappa^2 \langle Q\rangle$, and 
$\kappa^2 \langle Z\rangle$. However, the parameter $\kappa$ is not
known so that it should be factored out by taking the ratios of 
those observables. 
Consequently, there are four reconstructible observables:
\begin{eqnarray}
m_{\tilde{\chi}^\pm_1}, \qquad 
\sigma_{tot}(e^+e^-\rightarrow\tilde{\chi}^+_1\tilde{\chi}^-_1),\qquad
\frac{\langle P\rangle^2}{\langle Q\rangle}, \qquad 
   \frac{\langle Z\rangle}{\langle Q\rangle}.
\end{eqnarray}

Since the chargino-pair production is described by the four parameters, 
$\cos 2\phi_L$, $\cos 2\phi_R$, $\tan\beta$, and $m_{\tilde{\nu}}$ 
these four independent observables are expected to allow us to determine 
the SUSY parameters, $\tan\beta$, $M_2$ and $\mu$ (almost) completely.
In order to explicitly demonstrate the possibility of measuring the 
SUSY parameters, we make a case study with the assumption that the 
experimentally measured values of the four observables at 
$\sqrt{s}=200$ GeV are 
\begin{eqnarray}
m_{\tilde{\chi}^\pm_1}=90\ \ {\rm GeV}, \ \
\sigma_{tot}=2.5\ \ {\rm pb}, \ \
\frac{\langle P\rangle^2}{\langle Q\rangle}=-0.05\ \ {\rm pb},\ \
\frac{\langle Z\rangle}{\langle Q\rangle}=0.1.
\end{eqnarray}
For the ideal case with no experimental errors the observed values yield 
$m_{\tilde{\nu}}=197$ GeV, $\cos 2\phi_L=0.14$ and
$\cos 2\phi_R=0.03$.

\begin{figure}[htb]
\hbox to\textwidth{\hss\epsfig{file=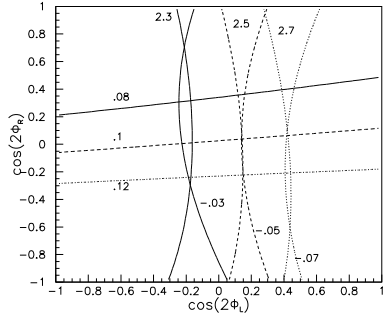,height=5cm}\hss}
\vskip 0.3cm
\caption{
A contour plot for $\sigma_{tot}=2.5\pm 0.2$ pb, 
$\langle P\rangle^2/\langle Q\rangle=-0.05\pm 0.02$ pb,
and $\langle Z\rangle/\langle Q\rangle=0.1\pm 0.02$ for 
$m_{\tilde{\chi}^\pm_1}=90$ GeV and $m_{\tilde{\nu}}=197$ GeV
at $\sqrt{s}=200$ GeV.}
\label{fig:unique}
\end{figure}

However, both the systematic and statistical errors should be considered
in real experimental situations. Also, the resolution power of the
observables constructed from angular correlations depends on the unknown 
parameter $\kappa$ rather strongly, and so its genuine estimates require 
prior understandings about the quantities such as the neutralino mass matrix 
and sfermion mass spectra. Fig.~\ref{fig:unique} illustrates how those 
errors may affect the determination of the cosines of two mixing angles under 
the assumption that
$\Delta\sigma_{tot}=\pm 0.2$ pb, 
$\Delta[\langle P\rangle^2/\langle Q\rangle]=\pm 0.02$ pb, and 
$\Delta[\langle Z\rangle/\langle Q\rangle]=\pm 0.02$ for $\sqrt{s}=200$ GeV
with a precisely measured chargino mass $m_{\tilde{\chi}^\pm}=90$ GeV.

From the measured values of $\cos 2\phi_L$, $\cos 2\phi_R$ and 
$m_{\tilde{\chi}^\pm_1}$, we can extract the allowed values for 
$\tan\beta$, $M_2$ and $\mu$. Defining  $x$ and $y$ satisfying 
$M_2=m_W(x+y)$ and $\mu=m_W(x-y)$, we find that there are four 
different solutions of $x$ and $y$
for give $\cos 2\phi_L$ and $\cos 2\phi_R$:
\begin{eqnarray}
&& (a)\ \ \left\{\begin{array}{l}
        x=\pm\cot(\phi_R-\phi_L)\sin\left(\beta-\frac{\pi}{4}\right), \\
        y=\pm\cot(\phi_R+\phi_L)\cos\left(\beta-\frac{\pi}{4}\right),
         \end{array}\right. \nonumber\\
&& (b)\ \ \left\{\begin{array}{l}
        x=\pm\cot(\phi_R+\phi_L)\sin\left(\beta-\frac{\pi}{4}\right), \\
        y=\pm\cot(\phi_R-\phi_L)\cos\left(\beta-\frac{\pi}{4}\right),
         \end{array}\right. 
\label{eq:x-y}
\end{eqnarray}
Depending on whether it is larger or smaller than the unity, 
the value of $\tan\beta$ is determined from $m_{\tilde{\chi}^\pm}$, 
$\cos 2\phi_L$ and $\cos 2\phi_R$, through the relations 
\begin{eqnarray}
&& \tan\beta =\left\{\begin{array}{l}
   \mp\frac{p^2-q^2\pm 2pq\sqrt{1+\eta}\sqrt{p^2+q^2-(1+\eta)p^2q^2}}{
         2(1+\eta)p^2q^2-(p-q)^2}\ \ {\rm for}\ \ \tan\beta\geq 1\\
   \mbox{ }\\
   \mp\frac{p^2-q^2\pm 2pq\sqrt{1+\eta}\sqrt{p^2+q^2-(1+\eta)p^2q^2}}{
          2(1+\eta)p^2q^2-(p+q)^2}\ \ {\rm for}\ \ \tan\beta\leq 1,
                    \end{array}\right.
\label{eq:tan_beta} 
\end{eqnarray}
where the overall $\mp$ is for (a) and (b) in Eq.~(\ref{eq:x-y}),
$\eta=m^2_{\tilde{\chi}^\pm_1}/m^2_W$,
$p^2-q^2=\sqrt{(1-\cos^2 2\phi_L)(1-\cos^2 2\phi_R)}$,
$pq=|\cos 2\phi_L-\cos 2\phi_R|/2$, and 
$p^2+q^2=1-\cos 2\phi_L \cos 2\phi_R$. 
Certainly only the $\tan\beta$ value satisfying each condition should be 
taken in Eq.~(\ref{eq:tan_beta}). To conclude, there exists at most a 
four-fold discrete ambiguity in determining $\tan\beta$, $M_2$, and $\mu$
in the pair production of lighter charginos at CERN LEP II.

\section*{Acknowledgments}

The author gratefully acknowledges the collaboration with A.~Djouadi,
J.~Kalinowski and P.M.~Zerwas.  
This work was supported by the KOSEF-DFG large collaboration
project, Project No. 96-0702-01-01-2.

\end{document}